\documentclass[nofootinbib,superscriptaddress,twocolumn,aps,prc,showpacs,10pt]{revtex4-1}
\usepackage{verbatim}
\usepackage{amsmath}
\usepackage{amssymb}
\usepackage{graphicx}
\usepackage{mathrsfs}
\usepackage{color}

\begin{document}

\title{Comment on the recent article "Fission dynamics of from saddle to scission and beyond" by Bulgac et al., published in Phys. Rev. {\bf C 100},  034615 (2019). 
}
\author{Sakir Ayik} \email{ayik@tntech.edu}
\affiliation{Physics Department, Tennessee Technological University, Cookeville, TN 38505, USA}

\author{Denis Lacroix} \email{lacroix@ipno.in2p3.fr}
\affiliation{Institut de Physique Nucl\'eaire, IN2P3-CNRS, Universit\'e Paris-Sud, Universit\'e Paris-Saclay, F-91406 Orsay Cedex, France}
\date{\today}

\begin{abstract}
In this note we provide answers to criticisms made the appendix D of the article \cite{Bul19} 
against to the theory developed recently and known a "Stochastic Mean-Field" (SMF) approach and its applications. 
Briefly, we list our replies and leave the judgment to the readers. 
\end{abstract}


\maketitle

In the article \cite{Bul19}, Bulgac and Coworkers made a number of remarks and critics on the "Stochastic Mean-Field" (SMF) approach \cite{Ayi08,Lac14}. We acknowledge
these critics and show below that most of them are based either on erroneous physics arguments or on miss-understanding of the SMF approach itself. The  SMF approach provides a description for incorporating the dynamical fluctuations in heavy-ion collisions beyond the mean-field at low energies where the dissipation and fluctuations are dominated by the one-body mechanism \cite{Ayi08,Lac14}.  
For this purpose, the SMF requires to generate an ensemble of mean-field events with the initial conditions specified by the quantal and thermal fluctuations in the initial state. 

This approach is already well documented, so we will not describe it in detail here. We would like to mention that the SMF theory is not meant to be an exact 
formulation of the many-body problem but an approximate Phase-Space method to incorporate beyond mean-field effects in transport theories. As an approximation, 
it cannot grasps all the many-body effects, but there are a number of indications that it can in many situations treat efficiently effects beyond the independent 
particle/quasi-particle picture. We mention for instance that:
\begin{itemize}
  \item As illustrated in \cite{Ayi08}, the SMF approach in the small amplitude limit becomes equivalent to the well established Balian-V\'en\'eroni [BV] description for the dispersion of one-body observables \cite{Bal84}.   
  \item It has been widely applied and validated against exact solutions in many situations \cite{Lac12,Lac13,Lac14b}. A careful analysis made in Ref. \cite{Lac12} have shown 
  that  it describes properly the dynamics in case of spontaneous symmetry breaking where the BV is not effective. 
  \item This theory was also the starting point for the various recent successful applications made on heavy-ion collisions \cite{Ayi09,Was09,Yil11,Yil14,Ayi15,Ayi16,Ayi17,Ayi18,Yil18,Ayi19}.  The SMF approach can determine the full distribution functions of the observables. 
  As illustrated in recent articles \cite{Ayi18,Ayi19}, the cross-section calculations for multi-nucleon transfer mechanism in the SMF approach gives very well description of data without any adjustable parameter, providing a strong support for the approach. It was also applied in Ref. \cite{Tan17} with some success. 
  \item We realized recently that the Phase-Space approach recently proposed in Ref. \cite{Dav17}, called fermionic-Truncated Wigner 
  Approximation (f-TWA), has direct connection with the SMF approach and gives an alternative independent formulation of the theory.   
\end{itemize}   

Despite these successes, the authors of Ref. \cite{Bul19} made strong criticisms on the SMF approach.     
We list below statements or comments made in Ref. \cite{Bul19} and reply to them one after the others:
\begin{enumerate}
  \item At several places in the article \cite{Bul19}, the authors state that the use of initial fluctuations only is contradictory with the Langevin approach where noise is continuously
  acting on trajectories. For instance, it is said in Ref. \cite{Bul19} (page 21):
  
  {\sl "In the stochastic mean-field model, fluctuations only stem from the fluctuations in the initial density [49] and the time evolution is exactly the usual time-dependent mean field. This ad hoc assumption is at odds with the Langevin approach..."   } 
  
 The incompatibility of initial fluctuations only with the Langevin approach is wrong. This is actually the opposite as it is well known in open quantum system (OQS) theory \cite{Bre02}. The Langevin force arrises from the absence of knowledge of the environment degrees of freedom (DOFs). This complexity of the bath/environment is
 treated as initial fluctuations  on its DOFs. Such fluctuations
 under certain conditions can lead to a random Markovian force acting continuously in time. A seminal example, where this is clearly illustrated is the Caldeira-Leggett model  \cite{Cal83}. Regarding the SMF approach, it is possible to project the microscopic description of SMF on a relevant collective sub-space by an adiabatic or geometric procedure as shown in \cite{Ayi08,Ayi18}.  The time evolution in the reduced collective subspace is governed by a generalized Langevin dynamics in which the stochastic force is determined by the fluctuations at the initial state. Such a description is fully consistent with the well-established Mori formalism \cite{Mor65}. 
  
  \item In Fig. 15 of Ref. \cite{Bul19}, the authors show that the diagonalisation of the one-body density leads to unphysical occupation numbers. This is indeed correct BUT  interpreting the event-by event one-body density obtained in the SMF approach as a quantum object does not make sense by itself. Indeed, the densities used in SMF should be interpreted as a statistical ensemble of one-body densities (treated as classical objects). The statistical properties of the initial densities are obtained by imposing that the average values of first and second moments equals the quantum average.  To be able to match statistical and quantal average, there is no other choice than exploring a wider class of one-body densities usually allowed 
  for Fermi systems. This is indeed (i) what is done in the SMF approach (ii) why the diagonalization of the one-body density, that implies to interpret it as a quantum object, 
is meaningless. Indeed, 
  in the SMF theory, only the average over events has a physical meaning. In our application, we always observed that the average one-body density has all required properties and leads after diagonalization to single-particle properties that are between $0$ and $1$. We would like to mention that the use of event-by-event unphysical one-body densities 
  to treat complex many-body effects is rather standard. Indeed, for instance, the authors mention \cite{Bul19} exact many-body methods (see for instance Eq. (1) of Ref. \cite{Bul19}). 
  In general, methods like auxiliary fields methods applied in real-time evolution, due to the necessity to have a term proportional to $\sqrt{i\Delta t}$ in the single-particle evolution, 
  the evolution is non-unitary and the one-body density is non-hermitian (see for instance \cite{Jul02,Lac05,Lac14}). Here also, it would not make sense to interpret a single trajectory. Still, the average evolution matches the exact evolution.

  \item Regarding the Eq. (D16) of Ref. \cite{Bul19} and the divergence of the local fluctuations in SMF. It is correct that the local fluctuations diverges when applying 
  the SMF technique. This divergence is actually not surprising because the fluctuations of the local density matches the fluctuations of the local one-body density 
  of a quasi-particle state that are also divergent\cite{Lan80}. Another example where the local density fluctuations diverges is simply the free gas \cite{Lan80}. So the problem is not 
  the SMF framework itself but the fact that the quantum fluctuations that are used  as reference are the ones of a quasi-particle vacuum. 
  
  It is worth mentioning that, starting from a pure Hamiltonian case, even using a contact interaction, the energy obtained by averaging the Hartree-Fock energy over events will match the energy of the initial state and no divergence of the energy will occur.  However, only when SMF approach is applied to a density functional theory (DFT), the special attention of terms like the one discussed in Eq. (D16) should be made. 
  
 In the SMF the initial fluctuations are specified so that the dispersion of one-body observables at the initial state matches with those calculated in the Hartree-Fock (HF) framework. This works fine for the macroscopic variables. The local density fluctuations have a singular behavior in the HF or HFB theory. As a result the fluctuations of the Energy Density Functional have a singular behavior as well. Physically, the local density fluctuations in volumes smaller than the minimum volume occupied by a single nucleon physically does not make sense. The criticisms of the appendix D are entirely based on this unphysical limit. For a correct description, a coarse-graining should be introduced in the local density fluctuations. The truncation of the particle-hole space in a narrow energy range around Fermi surface provides a possible for the coarse-graining of the local density fluctuations.  The application of the SMF to the spontaneous fission of  illustrates the simulation method with the truncation of the particle-hole space indeed works very well \cite{Tan17}. 

\item Last, the authors claims in appendix $D$ (page 21) that:

 {\sl "Since in the stochastic mean-field method fluctuations only stem from the fluctuations in the initial density [49] one would expect that their 
 conclusions should parallel ours..." } 

This is also incorrect. The fluctuations are different in the SMF approach compared to the one used in Ref. \cite{Bul19}. Indeed, in \cite{Bul19}, they select 
initial conditions on the potential energy landscape for instance shown in their Fig. 5 obtained using constrain HFB method with two collective degrees of freedom $Q_{20}$
and $Q_{30}$ and imposing specific symmetries. Although, most probably, other $Q_{\lambda 0}$ fluctuates from one initial condition to the other in \cite{Bul19}, some collective coordinates  
are frozen. This technique to get initial solutions, although rather common is rather specific. 
In the SMF approach, by treating fluctuations at the one-body level directly, all collective degrees of freedom $Q_{\lambda \mu}$ (including $\mu\neq 0$) are fluctuating 
initially without assuming a selection of a limited set of DOFs that are expected to be more relevant than the other ones. Another difference is the one-body density themselves that are 
allowed to explore a wider space compared to the one used in Ref. \cite{Bul19}. This absence of restriction in SMF  turns out to be crucial to grasps beyond mean-field effects.       
\end{enumerate}

\begin{acknowledgments}
S.A. gratefully acknowledges the IPN-Orsay for  the warm hospitality extended to him during his visits. 
This work is supported in part by the US DOE Grant No. DE-SC0015513. 
\end{acknowledgments}

\end{document}